\documentclass[conference]{IEEEtran}
\IEEEoverridecommandlockouts
\usepackage{cite}
\usepackage{amsmath,amssymb,amsfonts}
\usepackage{algorithmic}
\usepackage{graphicx}
\usepackage{textcomp}
\usepackage{graphicx} % Required for inserting images
\usepackage{amsmath,amsfonts,amssymb,mathrsfs}
\usepackage[scr=rsfso]{mathalfa}
\usepackage[table,xcdraw]{xcolor}
\usepackage{bbm}
\usepackage{verbatim}
\usepackage{pgfplots}
\pgfplotsset{compat=1.18}	
\usetikzlibrary{pgfplots.statistics} 
\usetikzlibrary{matrix}
\usepgfplotslibrary{groupplots}
\usepackage{tabularx}
\usepackage{multirow}
\usepackage{url}
\usepackage{array}
\usepackage{booktabs}
\usepackage{adjustbox}

\usepackage{color}

\def\BibTeX{{\rm B\kern-.05em{\sc i\kern-.025em b}\kern-.08em
    T\kern-.1667em\lower.7ex\hbox{E}\kern-.125emX}}
\begin{document}

\title{Whole-Body Image-to-Image Translation for a Virtual Scanner in a Healthcare Digital Twin}

\author{\IEEEauthorblockN{Anonymous Authors}}

\author{
    \IEEEauthorblockN{Valerio Guarrasi\IEEEauthorrefmark{1},
                      Francesco Di Feola\IEEEauthorrefmark{2},
                      Rebecca Restivo\IEEEauthorrefmark{1},
                      Lorenzo Tronchin\IEEEauthorrefmark{1}\IEEEauthorrefmark{3}, and
                      Paolo Soda\IEEEauthorrefmark{1}\IEEEauthorrefmark{2}}
    \IEEEauthorblockA{\IEEEauthorrefmark{1}Research Unit of Computer Systems and Bioinformatics, Department of Engineering, \\
    Università Campus Bio-Medico di Roma, Rome, Italy \\
    Emails: valerio.guarrasi@unicampus.it, rebecca.restivo@alcampus.it, p.soda@unicampus.it}
    \IEEEauthorblockA{\IEEEauthorrefmark{3}Department of Computing Science, Umeå University, Umeå, Sweden}
    \IEEEauthorblockA{\IEEEauthorrefmark{2}Department of Diagnostics and Intervention, Biomedical Engineering and Radiation Physics, Umeå University, Umeå, Sweden \\
    Email: francesco.feola@umu.se, paolo.soda@umu.se}
}

\maketitle

\begin{abstract}
Generating positron emission tomography (PET) images from computed tomography (CT) scans via deep learning offers a promising pathway to reduce radiation exposure and costs associated with PET imaging, improving patient care and accessibility to functional imaging. Whole-body image translation presents challenges due to anatomical heterogeneity, often limiting generalized models.
We propose a framework that segments whole-body CT images into four regions—head, trunk, arms, and legs—and uses district-specific Generative Adversarial Networks (GANs) for tailored CT-to-PET translation. Synthetic PET images from each region are stitched together to reconstruct the whole-body scan. Comparisons with a baseline non-segmented GAN and experiments with Pix2Pix and CycleGAN architectures tested paired and unpaired scenarios.
Quantitative evaluations at district, whole-body, and lesion levels demonstrated significant improvements with our district-specific GANs. Pix2Pix yielded superior metrics, ensuring precise, high-quality image synthesis. By addressing anatomical heterogeneity, this approach achieves state-of-the-art results in whole-body CT-to-PET translation.
This methodology supports healthcare Digital Twins by enabling accurate virtual PET scans from CT data, creating virtual imaging representations to monitor, predict, and optimize health outcomes.
\end{abstract}


\section{Introduction} \label{sec:Introduction}

Digital Twins (DTs) are a transformative healthcare innovation, offering digital replicas of physical objects, systems, or processes for real-time simulation and analysis~\cite{bib:grieves2017digital}. In medicine, DTs model patients, organs, or biological systems, enabling personalized care and predictive diagnostics~\cite{bib:bruynseels2018digital}. By integrating real-time data, DTs help monitor health, optimize treatments, and anticipate complications dynamically~\cite{bib:kamel2021digital}.

The concept of virtual scanning has recently emerged in radiology, enabling computational generation of medical images without physical scans~\cite{bib:wang2022development}. This approach leverages advancements in image-to-image translation, a field in deep learning that converts images from one domain to another while preserving key features~\cite{bib:isola2017image, bib:rofena2024deep}. In radiology, it enables transforming one imaging modality into another, such as converting CT to PET images, reducing patient exposure and scanning procedures~\cite{bib:dar2019image}.
Multimodality plays a critical role in radiology by integrating complementary information from different imaging techniques, enhancing diagnostic accuracy and clinical decision-making~\cite{bib:di2025graph,bib:guarrasi2024systematic,bib:guarrasi2023multi}. 
Generative Adversarial Networks (GANs) are widely employed for this purpose and have demonstrated success across various medical applications~\cite{bib:isola2017image}.
CT and PET are critical imaging modalities in clinical diagnostics, offering complementary information. CT scans provide high-resolution anatomical details using X-rays to detect structural abnormalities~\cite{bib:hsieh2003computed}, while PET scans visualize functional activity by detecting radiotracers like FDG, commonly used in oncology to identify malignant tumors~\cite{bib:gambhir2002molecular}. PET/CT, which combines these modalities, is the gold standard for tumor detection, staging, and monitoring by integrating metabolic and anatomical data~\cite{bib:beyer2000combined}.
However, PET/CT involves higher radiation doses due to combined CT and radiotracer exposure, raising safety concerns, particularly for repeated scans~\cite{bib:huang2009whole}. Additionally, it is costly and inaccessible in many regions due to the specialized equipment and radiotracers required~\cite{bib:marner2017clinical}. These limitations highlight the need for alternatives to obtain functional imaging without additional radiation or high costs.
Translating CT images into synthetic PET images using deep learning presents a promising solution, reducing radiation, lowering costs, and expanding access to functional imaging in resource-limited settings~\cite{bib:dayarathna2023deep}.

Whole-body image translation faces significant challenges due to the heterogeneity of anatomical structures and physiological functions across body regions. Variations in contrast, texture, and spatial resolution can adversely affect the performance of image-to-image translation models trained on the entire body~\cite{bib:dong2019synthetic}. Studies highlight that model effectiveness often varies by region, with some areas achieving higher translation accuracy than others due to these differences~\cite{bib:ge2019unpaired}.
For example, Dong et al.~\cite{bib:dong2019synthetic} reported more pronounced precision issues in specific body regions and recommended developing organ-specific networks tailored to each anatomical district. By capturing region-specific features, these models improve translation accuracy and reliability.
These findings emphasize the importance of specialized strategies that address regional variations in anatomy and imaging characteristics. Region-specific models are essential for advancing virtual scanning technologies and ensuring consistent accuracy and reliability across all anatomical areas.

A patient's DT is a virtual representation of their anatomy and physiology, built using patient-specific data to enable simulation, prediction, and clinical decision support.
In this context, the AI-based CT-to-PET translation exemplifies several core features of a DT by enriching anatomical data from CT with functional PET information, allowing clinicians to infer individual biological processes without additional invasive or costly imaging.
This approach simulates physiological processes by generating PET-equivalent data from CT scans, reflecting the DT’s core goal of replicating patient-specific metabolism and tissue characteristics in a virtual environment.
Furthermore, as PET imaging requires radioactive tracers, CT-to-PET translation offers a non-invasive alternative that reduces radiation exposure and facilitates repeated imaging when necessary.
Finally, these patient-specific, PET-like images enhance personalized clinical decision-making by highlighting areas of abnormal metabolic activity for diagnosis, tumor characterization, or treatment response assessment, aligning with the DT’s purpose of providing individualized insights over population-based norms.

To address the challenges of heterogeneity in whole-body image translation, we propose a novel approach that segments the body into distinct regions and employs district-specific GANs for image-to-image translation. Our methodology consists of three steps: (i) segmenting whole-body CT images into four districts—head, trunk, arms, and legs; (ii) processing each district independently using GANs tailored to its unique anatomical characteristics; (iii) stitching the district-specific PET images to reconstruct the whole-body PET scan.
By focusing on region-specific translation, this approach enhances the accuracy and reliability of synthetic PET images across all anatomical regions. We utilize two established GAN architectures, Pix2Pix~\cite{bib:isola2017image} and CycleGAN~\cite{bib:zhu2017unpaired}, due to their proven effectiveness in image-to-image translation tasks.
This method not only addresses performance variability across body regions but also demonstrates that specialized networks achieve superior results compared to single models trained on entire whole-body scans.

The remainder of this paper is organized as follows: %\bl Section~\ref{sec:Background} provides an overview of the background in virtual scanning; \bb
Section~\ref{sec:Materials} describes the dataset and preprocessing steps; Section~\ref{sec:Methods} elaborates on the proposed methodology and experimental setup; Section~\ref{sec:Results} presents the findings and discusses their implications; finally, Section~\ref{sec:Conclusions} summarizes the key insights and practical implications of the study.

\section{Materials} \label{sec:Materials}

\subsection{Dataset}

We utilized the FDG-PET/CT Lesions dataset~\cite{bib:gatidis2022whole}, a publicly available repository comprising 1,014 whole-body scans.
The dataset includes scans from patients diagnosed with malignant lymphoma, non-small cell lung cancer (NSCLC), malignant melanoma, as well as negative controls without PET-positive malignant lesions.
Specifically, 501 FDG-PET/CT scans of patients with malignant conditions and 513 scans of negative controls were collected between 2014 and 2018 at the University Hospital Tübingen.

The patient cohort ranged in age from 18 to 85 years, with a mean age of 52.3 years.
The dataset included 560 male and 454 female patients.
Inclusion criteria encompassed patients who underwent both PET and CT scans as part of their diagnostic workup for the specified conditions. 
Exclusion criteria included incomplete imaging data, significant motion artifacts, or any contraindications to PET/CT imaging procedures.
All imaging examinations were performed using a Siemens Biograph mCT PET/CT scanner.
The imaging protocol consisted of a diagnostic CT scan, primarily extending from the skull base to the mid-thigh level.
Intravenous contrast enhancement was administered in most cases, except for patients with contraindications.
CT acquisition parameters included a reference dose of 200 mAs, a tube voltage of 120 kV, and iterative reconstruction with a slice thickness of 2–3 mm.
For the PET component, whole-body FDG-PET scans were acquired 60 minutes after intravenous injection of 300–350 MBq of $^{18}$F-FDG. 
PET data were reconstructed using an ordered-subset expectation maximization algorithm with 21 subsets and 2 iterations, applying a Gaussian smoothing kernel of 2 mm and a matrix size of $400\times400$.

To facilitate model development and evaluation, the dataset was divided into a training set of 710 scans (70\% of the dataset), validation set of 101 scans (10\% of the dataset) and testing set of 203 scans (20\% of the dataset).
The splits were performed randomly but ensured that each subset maintained a balanced representation of the different diagnostic categories.
This stratification is crucial for the generalizability of our GAN models across various clinical scenarios.

A potential limitation of the dataset is the predominance of cases from a single institution, which may affect the generalizability of the model to data from other centers with different imaging protocols or patient populations.
Additionally, the dataset primarily includes scans from patients with specific oncological conditions, which may introduce a bias towards those pathologies and limit the model's performance when applied to other diseases or healthy populations.

%Despite these limitations, the FDG-PET/CT Lesions dataset was selected for its comprehensive coverage of whole-body PET/CT scans across a variety of oncological conditions and the inclusion of negative controls.
%This diversity enhances the robustness of our GAN models by providing a wide range of anatomical and pathological presentations.
%The availability of both PET and CT images for each patient enables the development of image-to-image translation models that can accurately synthesize PET images from CT inputs across different clinical scenarios. \bb

\subsection{Preprocessing}

Image preprocessing involved several steps to standardize and prepare the data for model training.
First, CT volumes were resampled to match the voxel size and spatial resolution of the corresponding PET images to ensure precise spatial alignment between modalities.
Resampling was performed using trilinear interpolation, providing a good balance between computational efficiency and interpolation accuracy.
Rigid registration was applied to correct for any remaining misalignments due to patient movement or differences in acquisition parameters.
Then, PET voxel values were converted to standardized uptake values (SUVs)~\cite{bib:boellaard2009standards}, which normalize tracer uptake by accounting for the injected dose of radiotracer and the patient's body weight.
Finally, to address variations in tracer uptake and enhance image contrast, a contrast adjustment was applied to the PET images, focusing on an SUV range of 0 to 20.
This range encompasses most clinically relevant uptake values while excluding extreme outliers that could skew the data.
Both CT and PET images were then normalized to a standard intensity range of $[0, 1]$ using min-max scaling, which maps the minimum and maximum voxel intensities to 0 and 1, respectively.
This normalization is essential for stabilizing the training of GAN models, as it ensures that the input and output data share a consistent scale.

\section{Methods} \label{sec:Methods}

\begin{figure*}[ht]
\centering
\includegraphics[width=\textwidth]{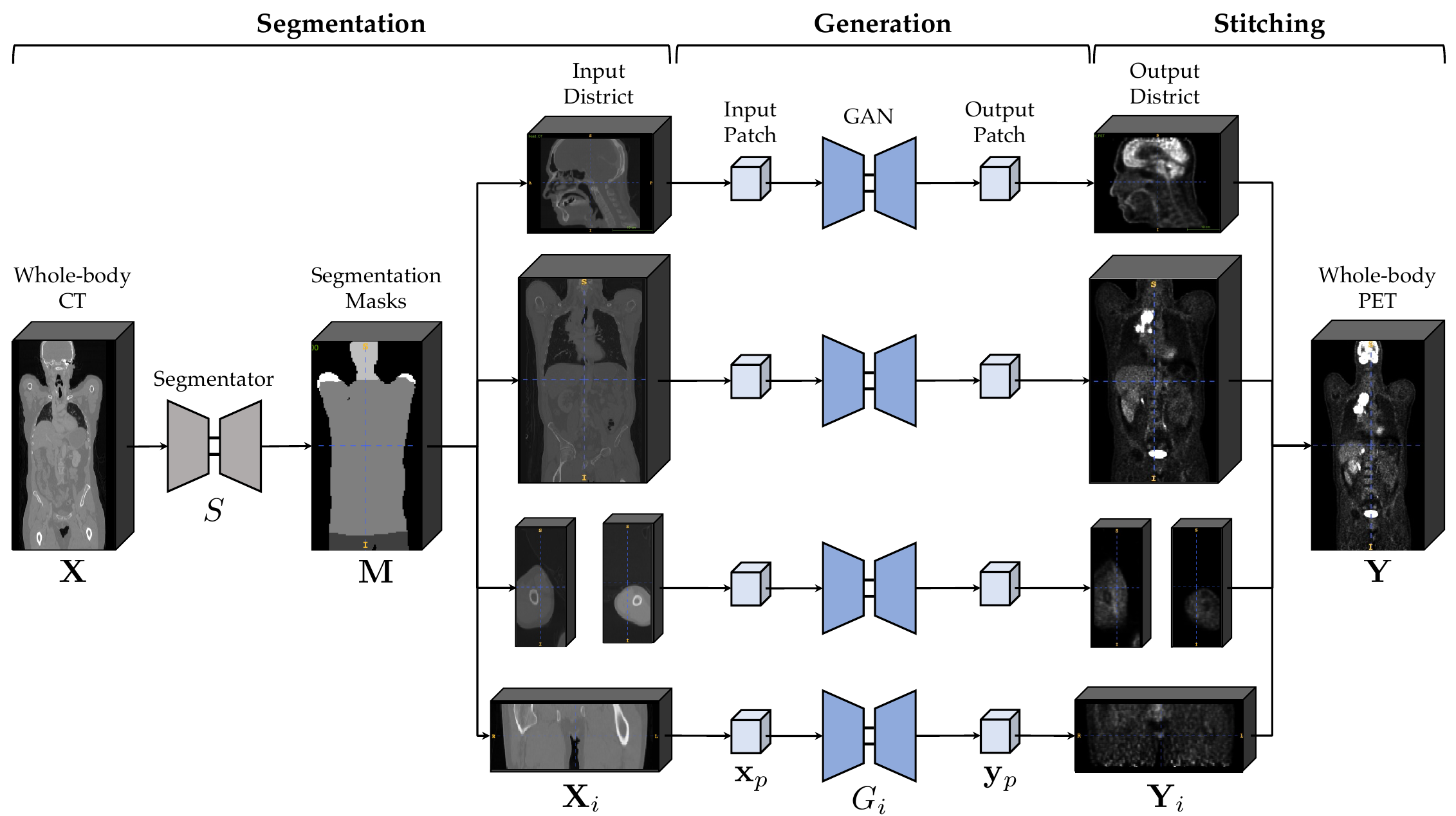}
\caption{Pipeline of the presented methodology.}
\label{fig:method}
\end{figure*}

In this study, we propose a novel methodology for synthesizing whole-body PET images from CT scans by utilizing district-specific GANs.
As shown in Figure~\ref{fig:method}, the approach comprises three primary stages: segmentation of whole-body CT images into distinct anatomical districts, training of individual GANs for each district to perform CT-to-PET translation, and stitching the generated district-specific PET images to reconstruct the final whole-body PET scan.
By decomposing the complex task of whole-body image translation into focused sub-tasks tailored to specific regions, we aim to improve the accuracy and reliability of synthesized images across all anatomical areas, thereby enhancing the clinical applicability of virtual scanning technologies.

Let $\mathbf{X} \in \mathbb{R}^{H \times W \times D}$ represent the input whole-body CT volume, where $H$, $W$, and $D$ denote the height, width, and depth dimensions, respectively.
The segmentation function $S$ partitions $\mathbf{X}$ into $N$ distinct anatomical districts:
\begin{equation}
\{ \mathbf{X}_i \}_{i=1}^N = S(\mathbf{X})
\end{equation}
such that:
\begin{equation}
\mathbf{X} = \bigcup_{i=1}^N \mathbf{X}_i
\end{equation}
and \( X_i \cap X_j = \emptyset \) for \( i \neq j \).
In our scenario $N=4$, corresponding to the head, trunk, arms, and legs.
For each district \( X_i \), a dedicated GAN model \( G_i \) is trained to learn the mapping from the CT domain to the PET domain:
\begin{equation}
G_i : \mathbf{X}_i \rightarrow  \mathbf{Y}_i
\end{equation}
with \( \hat{\mathbf{Y}}_i \) as the synthesized PET image corresponding to district \( \mathbf{X}_i \).
The final whole-body PET image \( \hat{\mathbf{Y}} \) is reconstructed by aggregating the generated district images:
\begin{equation}
\hat{\mathbf{Y}} = \bigcup_{i=1}^N \hat{\mathbf{Y}}_i
\end{equation}
In the following subsections, we detail each component of our methodology.

\subsection{Segmentation}

The initial phase of our methodology involves segmenting the whole-body CT scans into four major anatomical districts: head, trunk, arms, and legs.
We employed the MOOSE segmentator~\cite{bib:sundar2022fully}, an advanced semantic segmentation tool, to perform automated segmentation of the CT volumes.

Given $\mathbf{X}$, the input CT volume, let $\mathbf{M} = S(\mathbf{X})$ be the segmentation mask produced by applying the segmentation function $S$ to $\mathbf{X}$.
The mask $\mathbf{M} \in \{1, 2, 3, 4\}^{H \times W \times D}$ assigns each voxel $(x, y, z)$ a label corresponding to one of the four anatomical districts:
\begin{equation}
\mathbf{M}(x, y, z) =
\begin{cases}
    1, & \text{if voxel belongs to the head}, \\
    2, & \text{if voxel belongs to the trunk}, \\
    3, & \text{if voxel belongs to the arms}, \\
    4, & \text{if voxel belongs to the legs}.
\end{cases}
\end{equation}

For each district, we create a binary mask $\mathbf{M}_i$, where $i \in \{1, 2, 3, 4\}$, to isolate the specific region:
\begin{equation}
\mathbf{M}_i(x, y, z) =
\begin{cases}
    1, & \text{if } \mathbf{M}(x, y, z) = i \\
    0, & \text{otherwise}
\end{cases}
\end{equation}
Using these binary masks, we extract the corresponding district-specific CT volumes $\mathbf{X}_i$:
\begin{equation}
\mathbf{X}_i = \mathbf{X} \odot \mathbf{M}_i
\end{equation}
where $\odot$ denotes element-wise multiplication, effectively zeroing out voxels outside the $i$-th district.

Since the dataset comprises paired CT and PET images that are spatially aligned, we apply the same segmentation masks $\mathbf{M}_i$ to the PET volumes $\mathbf{Y}$ to obtain the district-specific PET images $\mathbf{Y}_i$:
\begin{equation}
\mathbf{Y}_i = \mathbf{Y} \odot \mathbf{M}_i
\end{equation}
where $\mathbf{Y} \in \mathbb{R}^{H \times W \times D}$ is the whole-body PET volume corresponding to $\mathbf{X}$.

\subsection{Generation}

In this phase, we train district-specific GANs to perform CT-to-PET translation for each anatomical district. We employed two GAN architectures: Pix2Pix~\cite{bib:isola2017image} and CycleGAN~\cite{bib:zhu2017unpaired}, implemented in their 3D versions to fully leverage the volumetric nature of medical imaging data, even if it poses significant computational challenges due to the large memory requirements.
The primary focus of our methodology is on the Pix2Pix model, which is well-suited for supervised learning scenarios with paired datasets, such as ours.
To demonstrate the extensibility of our pipeline to unpaired data scenarios, which are common in medical imaging due to difficulties in obtaining perfectly aligned datasets, we also performed experiments using the CycleGAN architecture.

For each district, we train a distinct GAN to learn the mapping from the CT domain to the PET domain:
\begin{equation}
G_i : \mathbf{X}_i \rightarrow \mathbf{Y}_i
\end{equation}

We utilized 3D convolutions to fully exploit the volumetric nature of medical imaging data.
To cope with the computational complexity posed by the 3D nature of our models, we adopted a patch-based training approach.
We divided the district-specific CT and PET volumes into smaller 3D patches of size $s \times s \times s$, where $s = 32$.
Let $\mathcal{P}_{\mathbf{X}_i}$ and $\mathcal{P}_{\mathbf{Y}_i}$ denote the sets of CT and PET training patches for the $i$-th district:
\begin{equation}
\mathcal{P}_{\mathbf{X}_i} = \{ \mathbf{x}_p \mid \mathbf{x}_p \subset \mathbf{X}_i, \mathbf{x}_p \in \mathbb{R}^{s \times s \times s} \}
\end{equation}
\begin{equation}
\mathcal{P}_{\mathbf{Y}_i} = \{ \mathbf{y}_p \mid \mathbf{y}_p \subset \mathbf{Y}_i, \mathbf{y}_p \in \mathbb{R}^{s \times s \times s} \}
\end{equation}
During training, we randomly extracted overlapping patches from $\mathbf{X}_i$ and $\mathbf{Y}_i$ and trained the corresponding GAN $G_i$ to minimize the respective loss function.

During the testing phase, we applied the trained $G_i$ to the CT volumes $\mathbf{X}_i$ to generate synthetic PET images.
To handle the high resolution and size of the volumes, we employed a sliding window technique with a window size of $s = 32$ and overlap $o = 16$.
We partitioned $\mathbf{X}_i$ into overlapping patches ${x}_p$ using a sliding window approach:
\begin{equation}
\mathbf{x}_p = \mathbf{X}_i(i:i+s-1, j:j+s-1, k:k+s-1)
\end{equation}
where $i, j, k$ are the starting indices along each axis, incremented by $s-o$.
Each CT patch $\mathbf{x}_p$ was input into $G_i$ to produce a synthetic PET patch $\hat{\mathbf{y}}_p$:
\begin{equation}
\hat{\mathbf{y}}_p = G_i(\mathbf{x}_p)
\end{equation}

\subsection{Stitching}

After generating the synthetic PET patches $\hat{\mathbf{y}}_p$, we reconstructed the full synthetic PET volume $\hat{\mathbf{Y}}_i$ for the $i$-th district by stitching the patches back together.
The overlapping regions between patches were averaged to ensure smooth transitions and reduce edge artifacts.

Let $\Omega_p$ denote the spatial coordinates of patch $\hat{\mathbf{y}}_p$ within $\hat{\mathbf{Y}}_i$. The value of each voxel $v$ in $\hat{\mathbf{Y}}_i$ is computed as:
\begin{equation}
\hat{\mathbf{Y}}_i(v) = \frac{\sum_{p: v \in \Omega_p} \hat{\mathbf{y}}_p(v)}{\sum_{p: v \in \Omega_p} 1}
\end{equation}
Once the synthetic PET volumes $\hat{\mathbf{Y}}_i$ for all districts are obtained, we assemble them to reconstruct the full synthetic whole-body PET image $\hat{\mathbf{Y}}$.
We initialize an empty volume $\hat{\mathbf{Y}} \in \mathbb{R}^{H \times W \times D}$.
For each district, we place the synthesized volume $\hat{\mathbf{Y}}_i$ into the corresponding location within $\hat{\mathbf{Y}}$:
\begin{equation}
\hat{\mathbf{Y}}(x, y, z) = 
\begin{cases} 
    \hat{\mathbf{Y}}_i(x, y, z), & \text{if } \mathbf{M}_i(x, y, z) = 1 \\
    0, & \text{otherwise}
\end{cases}
\end{equation}
By assembling the district-specific synthetic PET volumes into $\hat{\mathbf{Y}}$, we ensure anatomical correctness and spatial alignment with the original CT volume.
This process results in a coherent and continuous synthetic whole-body PET image that reflects the functional information corresponding to the input CT scan.

By combining segmentation, district-specific GAN generation, and systematic stitching, our methodology effectively addresses the challenges of anatomical heterogeneity in whole-body image translation.
This approach provides an accurate, reliable, and clinically applicable framework for synthesizing PET images from CT scans, with the potential to enhance virtual scanning technologies and extend to a wide range of clinical applications.

\section{Experimental Setup}

In this section, we describe the experimental setup used to evaluate proposed district specific virtual scanning methodology.
We outline a competitor approach in section, while we provide a comprehensive overview of the training configurations, computational resources, and evaluation metrics employed to ensure consistency and rigor across experiments. 

\subsection{Competitor Approach}

To establish a baseline for comparison and assess the effectiveness of our proposed district-specific GAN methodology, we implemented a standard whole-body GAN approach.
This conventional method aligns with current state-of-the-art practices in whole-body image-to-image translation~\cite{bib:dong2019synthetic}, where a single GAN model is trained to translate the entire whole-body CT scans into PET scans without any segmentation into anatomical districts.

In this approach, we trained both the Pix2Pix and CycleGAN architectures on the full whole-body images.
Given $\mathbf{X}$, the input whole-body CT volume, and $\mathbf{Y}$, the corresponding whole-body PET volume, the goal is to learn a mapping $G: \mathbf{X} \rightarrow \mathbf{Y}$ using a single GAN model that captures the complex relationships between CT and PET modalities across the entire body.

Similar to our district-specific GANs, we employed a 3D patch-based training strategy to handle the computational challenges associated with high-resolution volumetric data.

As before, the whole-body CT and PET volumes were partitioned into overlapping 3D patches of size $s \times s \times s$, where $s = 32$, with an overlap of $o = 16$ voxels along each dimension.
Formally, the set of training patches for the whole-body GAN is defined as:
\begin{equation}
\mathcal{P}_{\mathbf{X}} = \{ \mathbf{x}_p \mid \mathbf{x}_p \subset \mathbf{X}, \mathbf{x}_p \in \mathbb{R}^{s \times s \times s} \}
\end{equation}
\begin{equation}
\mathcal{P}_{\mathbf{Y}} = \{ \mathbf{y}_p \mid \mathbf{y}_p \subset \mathbf{Y}, \mathbf{y}_p \in \mathbb{R}^{s \times s \times s} \}
\end{equation}
Each patch $\mathbf{x}_p$ from the CT volume is paired with the corresponding patch $\mathbf{y}_p$ from the PET volume, and the GAN model is trained to minimize the loss function appropriate for the chosen architecture.

Comparing the results of the whole-body GANs with those of the district-specific GANs allows us to quantify the effectiveness of specializing models for different anatomical regions.

\subsection{Implementation Details}

The computational setup consisted of a high-performance computing cluster equipped with four NVIDIA A100 GPUs, each with 40 GB of memory, to accommodate the intensive computational requirements of 3D GAN training on volumetric medical imaging data.

For training the Pix2Pix and CycleGAN models, we utilized the Adam optimizer with hyperparameters set to $\beta_1 = 0.5$ and $\beta_2 = 0.999$ to promote stable convergence. The initial learning rate $\alpha$ was set to $2 \times 10^{-4}$.
A learning rate decay strategy was implemented, where the learning rate was linearly decreased to zero over the last 50 epochs of training to fine-tune the models.
Each model was trained for a total of 150 epochs, which was determined to be sufficient for convergence based on preliminary experiments.
Due to the memory constraints associated with processing 3D medical images, the batch size was set to 1 for both the generator and discriminator networks.
Gradient clipping was applied with a maximum norm of 5 to prevent exploding gradients during GAN training.
The hyperparameter $\lambda$ in the Pix2Pix loss function, balancing the adversarial loss and the $L_1$ reconstruction loss, was set to 100, as suggested by the original Pix2Pix implementation~\cite{bib:isola2017image} and validated through empirical testing.

Data augmentation techniques were employed to enhance the robustness of the models and prevent overfitting.
These included random rotations up to $\pm 10^\circ$ along each of the three spatial axes, random flipping along the axial, sagittal, and coronal planes, and adding Gaussian noise with a standard deviation $\sigma = 0.01$.

Each generator and discriminator was initialized with random weights and trained from scratch.
To ensure consistency across experiments, the same network architectures, hyperparameters, and training protocols were applied to both the district-specific GANs and the competitor whole-body GAN models.
The total training time for each district-specific GAN model was approximately 48 hours on the described computational setup.
In contrast, the whole-body GAN models required longer training times, averaging around 72 hours, due to the increased complexity of modeling the entire anatomical structure without regional specialization.

For evaluation, quantitative metrics, including Mean Absolute Error (MAE), Peak Signal-to-Noise Ratio (PSNR), and Structural Similarity Index Measure (SSIM), were computed on the same test set to assess the performance of the models.
All results were averaged over multiple runs to account for the stochastic nature of GAN training.

\section{Results and Discussions} \label{sec:Results}

In this section, we present the quantitative and qualitative evaluation of our proposed district-specific GAN approach for CT-to-PET image translation.
The results are analyzed at the district level, whole-body scans, and across different oncological conditions to demonstrate the advantages of anatomical specialization in GAN training, with a lesion-level evaluation to assess the clinical relevance of the generated synthetic PET images.
Finally, we discuss the implications of our findings for the development of digital twin frameworks in healthcare, highlighting the strengths and limitations of the proposed methodology and potential directions for future research.

\paragraph{District-Level and Whole-Body Evaluation}

\begin{figure}[h]
\centering
\begin{adjustbox}{width=\columnwidth}
\includegraphics{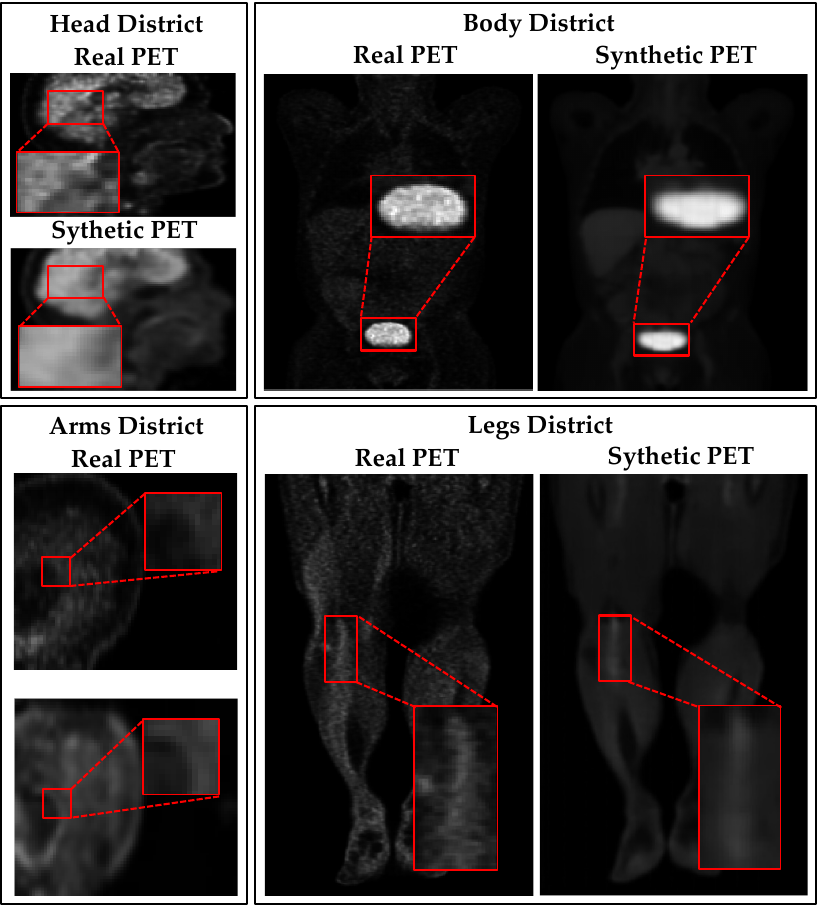}
\end{adjustbox}
\caption{Comparison of real and synthetic PET images across different anatomical districts.}
\label{fig:visual_results}
\end{figure}

We quantitatively evaluated the performance of our proposed district-specific GAN approach against the standard whole-body GAN competitor using three key metrics: MAE, PSNR, and SSIM.
These metrics were calculated for each anatomical district, i.e., head, trunk, arms, and legs, as well as for the whole-body scans.
For our method, the whole-body metrics were obtained by combining the outputs of the district-specific networks.
The results, summarized in Table~\ref{tab:district}, report the mean and standard error of each metric across all test patients.
Our proposed district-specific GAN approach, utilizing both the Pix2Pix and CycleGAN architectures, consistently outperformed the competitor whole-body GAN method across all anatomical regions and evaluation metrics.
Our proposed method exhibited lower MAE values across all districts, indicating a closer approximation to the ground truth PET images.
Similarly, higher PSNR values were observed with our method, reflecting better reconstruction quality and reduced noise levels.
The SSIM scores were also higher for our approach, suggesting superior preservation of structural information and image textures.
Moreover, when comparing the two architectures within our method, the Pix2Pix models generally outperformed the CycleGAN models.
This outcome can be attributed to the supervised nature of Pix2Pix, which benefits from paired training data, allowing for more precise learning of the mapping between CT and PET images.
However, both architectures in our method surpassed their counterparts in the competitor approach, underscoring the advantage of district-specific training.
\figurename~\ref{fig:visual_results} shows a comparison of real and synthetic PET images across different anatomical districts, demonstrating the effectiveness of our approach in synthesizing PET from CT.
Zoomed-in Regions of Interest, focus on key areas of metabolic activation, where we can note that the majority of activated regions are accurately preserved in synthetic PET images.

\begin{table}[h!]
    \centering
    \caption{District and Whole-Body Level Performance.}
    \label{tab:district}
    \begin{adjustbox}{width=\columnwidth}
    \begin{tabular}{|c|c|cc|cc|}
        \hline
        \multirow{2}{*}{\textbf{District}} & \multirow{2}{*}{\textbf{Metric}} & \multicolumn{2}{c|}{\textbf{Proposed District-Specific GAN}} & \multicolumn{2}{c|}{\textbf{Competitor Whole-Body GAN}} \\
        \cline{3-6}
        & & \textbf{Pix2Pix} & \textbf{CycleGAN} & \textbf{Pix2Pix} & \textbf{CycleGAN} \\ 
        \hline
        \multirow{3}{*}{\textbf{Head}} 
        & MAE $\downarrow$ & \textbf{0.023} ± 0.005 & 0.025 ± 0.005 & 0.044 ± 0.064 & 0.050 ± 0.070 \\
        & PSNR $\uparrow$ & \textbf{26.16} ± 2.18 & 24.78 ± 1.82 & 25.88 ± 7.60 & 23.50 ± 8.00 \\
        & SSIM $\uparrow$ & 0.76 ± 0.03 & \textbf{0.78} ± 0.04 & 0.65 ± 0.23 & 0.60 ± 0.25 \\
        \hline
        \multirow{3}{*}{\textbf{Trunk}} 
        & MAE $\downarrow$ & \textbf{0.008} ± 0.001 & 0.022 ± 0.001 & 0.018 ± 0.016 & 0.020 ± 0.018 \\
        & PSNR $\uparrow$ & \textbf{32.36} ± 1.78 & 27.63 ± 1.13 & 30.63 ± 2.89 & 28.00 ± 3.00 \\
        & SSIM $\uparrow$ & \textbf{0.87} ± 0.02 & 0.59 ± 0.04 & 0.70 ± 0.16 & 0.65 ± 0.18 \\
        \hline
        \multirow{3}{*}{\textbf{Arms}} 
        & MAE $\downarrow$ & \textbf{0.005} ± 0.002 & 0.008 ± 0.003 & 0.010 ± 0.001 & 0.012 ± 0.002 \\
        & PSNR $\uparrow$ & \textbf{40.28} ± 3.42 & 36.38 ± 3.55 & 31.65 ± 0.23 & 30.00 ± 0.50 \\
        & SSIM $\uparrow$ & \textbf{0.94} ± 0.03 & 0.84 ± 0.04 & 0.83 ± 0.03 & 0.80 ± 0.05 \\
        \hline
        \multirow{3}{*}{\textbf{Legs}} 
        & MAE $\downarrow$ & \textbf{0.010} ± 0.002 & \textbf{0.010} ± 0.002 & 0.013 ± 0.005 & 0.015 ± 0.006 \\
        & PSNR $\uparrow$ & \textbf{35.68} ± 2.50 & 34.34 ± 1.50 & 31.42 ± 0.80 & 30.50 ± 1.00 \\
        & SSIM $\uparrow$ & 0.76 ± 0.04 & \textbf{0.79} ± 0.04 & 0.76 ± 0.10 & 0.74 ± 0.12 \\
        \hline
        \multirow{3}{*}{\textbf{Whole-Body}} 
        & MAE $\downarrow$ & \textbf{0.011} ± 0.006 & 0.014 ± 0.001 & 0.020 ± 0.005 & 0.022 ± 0.006 \\
        & PSNR $\uparrow$ & \textbf{31.43} ± 1.00 & 29.10 ± 0.86 & 28.50 ± 1.50 & 27.00 ± 2.00 \\
        & SSIM $\uparrow$ & \textbf{0.76} ± 0.01 & 0.56 ± 0.02 & 0.68 ± 0.05 & 0.65 ± 0.06 \\
        \hline
    \end{tabular}
    \end{adjustbox}
\end{table}

\paragraph{Lesion-Level Evaluation}

Given the clinical importance of accurately detecting and representing lesions in PET images for diagnostic purposes, we conducted an additional evaluation focusing specifically on lesion regions.
Using manually annotated lesion masks from the FDG-PET/CT Lesions dataset, we calculated MAE, PSNR, and SSIM metrics exclusively over the lesion voxels. 
The lesion-level results, presented in Table~\ref{tab:lesion}, indicate that our district-specific GAN approach significantly outperformed the competitor in synthesizing PET activity within lesion regions.
Our proposed approach model achieved the lowest MAE in lesion regions, indicating a more accurate representation of tracer uptake in lesions compared to the competitor approach model.
Similarly, higher PSNR and SSIM values were observed with our method, suggesting improved image quality and better preservation of structural details within lesions.
As before, when comparing the two architectures within our method, the Pix2Pix models generally outperforms the CycleGAN models.

\begin{table}[h!]
    \centering
    \caption{Lesion-Level Performance.}
    \label{tab:lesion}
    \begin{adjustbox}{width=\columnwidth}
    \begin{tabular}{|c|cc|cc|}
        \hline
        \multirow{2}{*}{\textbf{Metric}} & \multicolumn{2}{c|}{\textbf{Proposed District-Specific GAN}} & \multicolumn{2}{c|}{\textbf{Competitor Whole-Body GAN}} \\
        \cline{2-5}
        & \textbf{Pix2Pix} & \textbf{CycleGAN} & \textbf{Pix2Pix} & \textbf{CycleGAN} \\
        \hline
        MAE $\downarrow$ & \textbf{0.012} ± 0.003 & 0.015 ± 0.004 & 0.020 ± 0.005 & 0.025 ± 0.006 \\
        PSNR $\uparrow$ & \textbf{30.50} ± 1.20 & 28.80 ± 1.50 & 27.00 ± 1.80 & 25.50 ± 2.00 \\
        SSIM $\uparrow$ & \textbf{0.80} ± 0.02 & 0.75 ± 0.03 & 0.68 ± 0.04 & 0.62 ± 0.05 \\
        \hline
    \end{tabular}
    \end{adjustbox}
\end{table}

\paragraph{Oncological Condition Level Evaluation}

To further assess the robustness and clinical applicability of our proposed district-specific GAN approach, we conducted an evaluation based on different oncological conditions present in our dataset, i.e., malignant lymphoma, NSCLC, malignant melanoma, as well as negative controls.
Evaluating our method across these conditions allows us to determine its effectiveness in capturing the varying metabolic activities associated with different types of cancer.
The quantitative results are presented in Table~\ref{tab:oncology}, where for each oncological condition, we grouped the test patients accordingly and computed the evaluation metrics.
The metrics were calculated over the entire PET volume for each patient, providing a comprehensive assessment of image quality and accuracy within each condition.
Our proposed district-specific GAN approach consistently outperformed the competitor across all oncological conditions, as evidenced by lower MAE values and higher PSNR and SSIM scores.
The enhanced performance of our proposed method across different oncological conditions highlights its potential clinical utility.
Furthermore, we observe that the Negative Controls exhibit slightly better performance, which aligns with the original data distribution and reflects the lower complexity of this condition.

\begin{table}[h!]
\centering
\renewcommand{\arraystretch}{1.5}
\caption{Oncological Condition Level Performance.}
\label{tab:oncology}
\begin{adjustbox}{width=\columnwidth}
\begin{tabular}{|c|c|cc|cc|}
\hline
\multirow{2}{*}{\textbf{Condition}} & \multirow{2}{*}{\textbf{Metric}} & \multicolumn{2}{c|}{\textbf{Proposed District-Specific GAN}} & \multicolumn{2}{c|}{\textbf{Competitor Whole-Body GAN}} \\ 
\cline{3-6}
 &  & \textbf{Pix2Pix} & \textbf{CycleGAN} & \textbf{Pix2Pix} & \textbf{CycleGAN} \\ 
\hline
\multirow{3}{*}{\textbf{Lymphoma}} 
 & MAE $\downarrow$ & \textbf{0.012} ± 0.003 & 0.015 ± 0.004 & 0.020 ± 0.005 & 0.022 ± 0.006 \\ 
 & PSNR $\uparrow$ & \textbf{31.00} ± 1.20 & 29.00 ± 1.50 & 28.00 ± 1.80 & 27.00 ± 2.00 \\ 
 & SSIM $\uparrow$ & \textbf{0.75} ± 0.02 & 0.70 ± 0.03 & 0.68 ± 0.04 & 0.65 ± 0.05 \\ 
\hline
\multirow{3}{*}{\textbf{NSCLC}} 
 & MAE $\downarrow$ & \textbf{0.013} ± 0.003 & 0.016 ± 0.004 & 0.021 ± 0.005 & 0.023 ± 0.006 \\ 
 & PSNR $\uparrow$ & \textbf{30.50} ± 1.30 & 28.50 ± 1.60 & 27.50 ± 1.80 & 26.50 ± 2.00 \\ 
 & SSIM $\uparrow$ & \textbf{0.74} ± 0.02 & 0.69 ± 0.03 & 0.66 ± 0.04 & 0.63 ± 0.05 \\ 
\hline
\multirow{3}{*}{\textbf{Melanoma}} 
 & MAE $\downarrow$ & \textbf{0.011} ± 0.002 & 0.014 ± 0.003 & 0.019 ± 0.004 & 0.021 ± 0.005 \\ 
 & PSNR $\uparrow$ & \textbf{31.50} ± 1.10 & 29.50 ± 1.40 & 28.50 ± 1.60 & 27.50 ± 1.80 \\ 
 & SSIM $\uparrow$ & \textbf{0.76} ± 0.02 & 0.71 ± 0.03 & 0.69 ± 0.04 & 0.66 ± 0.05 \\ 
\hline
\multirow{3}{*}{\textbf{Negative Controls}} 
 & MAE $\downarrow$ & \textbf{0.009} ± 0.002 & 0.012 ± 0.003 & 0.017 ± 0.004 & 0.019 ± 0.005 \\ 
 & PSNR $\uparrow$ & \textbf{32.50} ± 1.00 & 30.50 ± 1.20 & 29.50 ± 1.40 & 28.50 ± 1.60 \\ 
 & SSIM $\uparrow$ & \textbf{0.78} ± 0.01 & 0.73 ± 0.02 & 0.71 ± 0.03 & 0.68 ± 0.04 \\ 
\hline
\end{tabular}
\end{adjustbox}
\end{table}

Statistical analysis using paired two-tailed Student's $t$-tests confirmed that the performance improvements of our proposed method were statistically significant ($p < 0.01$) across all metrics, regions, and conditions, including district level, whole-body scans, lesion-level evaluations, and various oncological conditions; indicating the robustness of our approach in handling different anatomical regions and types of cancer.

\section{Conclusions} \label{sec:Conclusions}

In this study, we introduced a novel methodology for synthesizing whole-body PET images from CT scans using district-specific GANs. 
By segmenting the body into four primary anatomical districts, i.e., head, trunk, arms, and legs, we trained specialized GANs tailored to each region's unique anatomical and functional characteristics.
This approach effectively addressed the challenges posed by anatomical heterogeneity in whole-body imaging, demonstrating that the district-specific GANs significantly outperformed the standard whole-body GAN models across the evaluated metrics.

Our work has the potential to substantially impact clinical practice and medical imaging, particularly within DT paradigm in healthcare.
By enabling accurate virtual PET scanning from existing CT data, our approach eliminates the need for additional physical scans, reducing reliance on conventional PET imaging.
This lowers healthcare costs, minimizes patient exposure to ionizing radiation, especially for those requiring multiple scans, and expands access to functional imaging in medical centers without PET facilities~\cite{bib:caruso2024deep}.

Future work will refine the segmentation process by including more detailed subdivisions or organ-specific segmentation, potentially enabling even greater specialization of the GANs.
Additionally, we plan to integrate explainable artificial intelligence techniques to enhance transparency and trust in the GAN-generated PET images, particularly for critical clinical applications~\cite{bib:guarrasi2024multimodal}.
Furthermore, we aim to evaluate the generalizability of our approach across diverse datasets to ensure robustness and adaptability in varying clinical settings~\cite{bib:ruffini2024multi,bib:guarrasi2022optimized,bib:mogensen2025optimized}.

\begin{comment}
\section*{Acknowledgment}
Resources are provided by the National Academic Infrastructure for Supercomputing in Sweden (NAISS) and the Swedish National Infrastructure for Computing (SNIC) at Alvis @ C3SE, partially funded by the Swedish Research Council through grant agreements no. 2022-06725 and no. 2018-05973.
This work was partially founded by: i) PNRR MUR project PE0000013-FAIR, ii)  Cancerforskningsfonden Norrland project MP23-1122, iii) Kempe Foundation project JCSMK24-0094, iv) PNRR M6/C2 project PNRR-MCNT2-2023-12377755, v) Università Campus Bio-Medico di Roma under the program ``University Strategic Projects'' within the project ``AI-powered Digital Twin for next-generation lung cancEr cAre (IDEA)''.
\end{comment}

\bibliographystyle{IEEEtran}
\bibliography{mybibfile_etal}

\end{document}